\documentclass[pra,twocolumn,floatfix,showpacs]{revtex4}

\usepackage{epsfig}

\newcommand{\be}[1]{\begin{equation}\label{#1}}
\newcommand{\ee}{\end{equation}}   
\newcommand{\bea}{\begin{eqnarray}}
\newcommand{\eea}{\end{eqnarray}} 
\newcommand{\eq}[1]{(\ref{#1})}
\newcommand{\tbl}[1]{table~\ref{#1}}

\newcommand{\ba}[1]{\begin{array}{#1}}   
\newcommand{\ea}{\end{array}}   

\newcommand{\non}{\mbox{\scriptsize non}}	
\newcommand{\etal}{\emph{etal}~}	

\begin{document}

\title{A mapping approach to synchronization in the "Zajfman trap":
stability conditions and the synchronization mechanism}

\author{Tiham\'{e}r\ Geyer and David\ J\ Tannor}
\affiliation{Department of Chemical Physics, Weizmann Institute of
Science, Rehovot 76100, Israel} 


\begin{abstract}
	We present a two particle model to explain the mechanism that
	stabilizes a bunch of positively charged ions in an "ion trap
	resonator" [Pedersen \emph{etal}, \emph{Phys. Rev. Lett.}
	\textbf{87} (2001) 055001]. The model decomposes the motion of the
	two ions into two mappings for the free motion in different parts
	of the trap and one for a compressing momentum kick. The ions'
	interaction is modelled by a time delay, which then changes the
	balance between adjacent momentum kicks. Through these mappings we
	identify the microscopic process that is responsible for
	synchronization and give the conditions for that regime.
\end{abstract}

\maketitle

\section{Introduction}

Ion traps are widely used to store ions for long periods to obtain
high resolution spectroscopy. Ions of the same charge repel each other
through their long range Coulomb potential, so the trap has the
purpose of localizing the ions in the measurement region. Recently a
quite surprising behavior was discovered \cite{PED01}, where, under
special conditions, ions in an ion trap resonator do not diffuse into
the whole trap but stay together as a bunch for arbitrarily long
times. This synchronized, collective motion occurs only for certain
parameters of the trap fields, but in these regions it is stable.

This observation, which is puzzling and of intriguing scientific
interest in its own right, has important technological applications.
Pedersen \emph{etal}~ have suggested the use of the small "table top"
ion trap as a time--of--flight mass spectrometer. As the observation
time, and therefore the effective length of the spectrometer, are in
principle only limited by collisions between the ions, mass
resolutions can be envisioned which are otherwise only achieved in
storage rings \cite{MAR98}. Indeed, the high resolution spectroscopy
suggested in \cite{PED01} has now been achieved \cite{STR02}.

Until now this synchronization effect has not been fully understood;
from the experimental observations \cite{PED02a, PED02b} it appears
that the trap has to be operated in a regime where the ions' periods
in the trap \emph{increase} with their energy. Other tests point out
that the focussing of the beam inside the mirrors is important.
Numerical simulations confirm these empirical findings, but they, too,
can not decide, if the requirements found so far, are complete and if
they really stem from the underlying microscopic process.

A macroscopic explanation of the synchronization effect in terms of a
"negative mass instability" was recently presented in \cite{STR02}. It
confirms that the ions' period has to increase with their energy and
shows that a minimal density inside the bunch is necessary to support
synchronization. But, being a mean field treatment, it can not give a
detailed explanation of the underlying microscopic dynamics and it is
insensitive to certain properties of the trapping field.

The aim of this paper is therefore, to set up a complementary
microscopic model of the ions in the trap, which is simple enough to
be understood completely. With this model we then can explain the
basic mechanism, determine the necessary conditions for
synchronization, check their completeness and finally understand how
the size of the bunch depends on the parameters of the trap and those
of the ion beam.

In this paper we concentrate on deriving the conditions for stability
and we explain and illustrate the fundamental mechanism. The stability
limits and the behavior of the macroscopic bunch will be presented in
a following paper \cite{GEY02}.

This paper is organized as follows: In section \ref{sec:ModelOfTrap}
we explain the model of the trap. In section
\ref{sec:TwoNonInteracting} we lay out the framework of the dynamics
in terms of the mappings for two non interacting ions. Then, in
section \ref{sec:AddingInteraction}, we explain how their interaction
is incorporated in the mappings. With the interaction added we derive
the conditions for bunching, section \ref{sec:LinearStability}, and
explain the underlying mechanism in section \ref{sec:Explaining}. In
section \ref{sec:Numericals} we confirm that the conditions and
explanations given are in fact applicable to the experiment.

\section{The model of the trap}
\label{sec:ModelOfTrap}

The experimental setup and the observed behavior is described in great
detail in \cite{PED02a, PED02b}. We will therefore only cite what we
need to build up the model. In the experiment a bunch of ions is
injected into the "ion trap resonator" and its width observed, when it
passes through a ring shaped pickup electrode in the trap's center.
When the field gradient in the electrostatic mirrors is below a
certain threshold, the bunch does not diffuse. 

\epsfxsize=6cm
\epsfysize=3.375cm
\begin{figure}[t]
	\epsffile{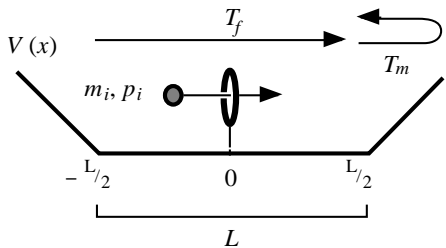}
	\caption{Model of the trap: the simplified potential $V(x)$ 
	\eq{eq:trapPotential} with the ions taking time $T_f$ through the 
	central part and $T_m$ to return from the mirror. The ring shaped 
	pickup electrode is located in the center of the trap.}
	\label{fig:TrapGeom}
\end{figure}
Numerical simulations with many ions performed by Pedersen \etal
have shown synchronization in a one dimensional model \cite{PED02b}. We
will consider a one dimensional model, too, and use the same 
simplified trap potential, which
consists of
the central field free region of length $L$ and two slopes with a
constant gradient $F$, the electrostatic mirrors. The two einzel
lenses, which focus the beam into the mirrors \cite{PED02a}, are
neglected. They are only necessary for keeping the beam inside the 
trap. This setup is modelled by the potential
\be{eq:trapPotential}
	V(x) = \left\{ 
	\begin{array}{ccl}
		0                     & \mbox{ when } & |x| \leq \frac{L}{2}  
			\\[0.2cm]
		(|x|-\frac{L}{2})\, F & \mbox{ when } & |x| > \frac{L}{2}
	\end{array}
	\right.
\ee

We will not treat the whole bunch, but will look at the behavior of
only two identical ions in the trap and explain how these two ions
synchronize their motion.

Since synchronization is a property of the relative coordinate, we
describe these two ions in relative and center of mass (CM)
coordinates:
\bea
	\label{eq:CMRelReduced}
	\begin{array}{ccccc}
		M = 2m & \quad & R = \frac{x_1 + x_2}{2} & \quad & 
			P = p_1 + p_2  \\[0.2cm]
		\mu = \frac{m}{2} & \quad & x = x_1 - x_2 & \quad & 
			p = \frac{p_1 - p_2}{2}
	\end{array}
\eea
%
Capital letters denote CM properties and lowercase letters are used
for the relative coordinate. With the ion--ion interaction $W(x)$ the
Hamiltonian now reads:
\be{eq:Hamiltonian}
	H = \frac{P^2}{2M} + \frac{p^2}{2\mu} + 
		qV(R+\textstyle{\frac{x}{2}}) + qV(R-\textstyle{\frac{x}{2}}) 
		+ q^2 W(x)
\ee
Without loss of generality we set the charge of the ions to $q=+1$ in 
the following.
Though $W(x)$ could be an arbitrary two body interaction we will think
of it as the repulsive Coulomb potential between two positively
charged ions.

Our model will not be exact. In order to check the validity of the
various approximations which we perform, we insert the values from the
original experiment, in which the bunching effect was observed
initially \cite{PED01}. These values are given in
\tbl{tbl:Dimensions}. If not otherwise stated we use atomic units in
the following.
\begin{table}[t]
	\centering
	\begin{tabular}{|c|ccc|}
		\hline
		quantity & value & & in a.u.  \\
		\hline
		\hline
		length $L$ & 200 mm & = & $3.78 \times 10^9$ a.u. \\
		mirror field gradient $qF$ & 80 $\frac{keV}{m}$ & = &
			$1.56\times 10^{-7}$ $\frac{a.u.}{a.u}$  \\
		ion mass $m$ (Ar$^+$)& 40 amu & = & $7.35 \times 10^4$ a.u. \\
		ion energy $E_0$ & 4.2 keV & = & $154$ a.u. \\
		ion momentum $p_i$ & & & $4.76 \times 10^3$ a.u. \\
		beam radius $d$ & 10 $\mu$m & = & $1.89 \times 10^5$ a.u. \\ 
		\hline
		time in the mirror $T_m$ & 1.48 $\mu$s & = & $6.10 \times 10^{10}$ a.u. \\
		time in the center $T_f$ & 1.41 $\mu$s & = & $5.84 \times 10^{10}$ a.u. \\
 		dispersion parameter $\alpha$ & & 0.956 & \\
		\hline
	\end{tabular}
	\caption{Parameters and sizes used in our calculations. These
	values are the same as in the original experiment, in which
	bunching was originally discovered \cite{PED01}.}
	\label{tbl:Dimensions}
\end{table}

An ion with the momentum $p_0 = \sqrt{2mE_0}$ will spend time $T_m$ in
the mirror, which can also be expressed in terms of the CM momentum:
\be{eq:TimeInTheMirror}
	T_m = \frac{2p_0}{F} = \frac{P}{F}
\ee
Ions with a velocity $p_0/m = P/M$ need the time $T_f$ to pass
through the central field free region of the trap:
\be{eq:TimeInTheFlat}
	T_f = \frac{Lm}{p_0} = \frac{LM}{P} := \alpha T_m
\ee
Here we have introduced the ``dispersion parameter'' $\alpha =
\frac{LmF}{2p_0^2}$ as the ratio of the two times $T_f$ and $T_m$.
With these the total time $T$ for the period of one ion becomes:
\be{eq:TimeForPeriod}
	T = 2T_m + 2T_f = 2(1+\alpha)T_m = \frac{4p_0}{F} + \frac{2Lm}{p_0}
\ee
From this equation we calculate the "dispersion" $\frac{\partial
T}{\partial p_0}$ of the trap:
\be{eq:Dispersion}
	\frac{\partial T}{\partial p_0} = \frac{4}{F} - \frac{2Lm}{p_0^2}
	= \frac{4}{F}(1-\alpha)
\ee
In the experiment and in the theoretical model of reference
\cite{STR02} the ions were found to synchronize, when the dispersion
is positive, i.e, when an ion with a higher momentum (energy) takes
longer to complete one period in the trap. With \eq{eq:TimeInTheFlat}
we have $\alpha <1$: the ions then spend more time in the mirror than
in the central region.

Two ions with different momenta have different periods, which in turn
can be reformulated as a difference $\Delta x$ in their distance when we
propagate both ions for the same time $T$. To calculate $\Delta x$, we
linearize $T$ \eq{eq:TimeForPeriod} around the mean momentum
$\frac{P}{2}$. With the difference in the ions' momenta of $\Delta p =
2p$ \eq{eq:CMRelReduced} the difference between their periods is
$\Delta T = \frac{\partial T}{\partial p_0} \Delta p$. During that
interval the ions move with the velocity $\frac{P}{M}$, leading to an
increase in separation of
\bea
	\Delta x(p) & = & \frac{P}{M} \left(
		\left.\frac{\partial T}{\partial p_0}\right| _{P/2} 
		\Delta p\right)  \nonumber \\
	& = & \frac{2 T_m}{\mu} (1 - \alpha) p
	\label{eq:IncreaseDistDisp}
\eea
%
Here we have used \eq{eq:TimeInTheMirror}---\eq{eq:Dispersion}. Again
$\alpha$ appears as the central quantity, determining the sign of
$\Delta x$.

In \cite{PED01} the energy spread of the ion source is given as 
$\frac{\Delta v}{v} < 0.1\% $, which translates into a relative 
momentum of $p < 4$ a.u. With the parameters of \tbl{tbl:Dimensions} 
the relative error between the linearized $\Delta T$ and the correct 
difference $T(\frac{P}{2}+p) - T(\frac{P}{2}-p)$ is well below 
$\mathcal{O}(10^{-5})$.

Pedersen \etal defined a parameter $\alpha$, too, which we refer to as
$\alpha_p$ \cite{PED02b}. From its definition as $\alpha_p =
\frac{1}{T}\frac{\partial T}{\partial E_0}$ we calculate with
\eq{eq:TimeInTheMirror}--\eq{eq:Dispersion} the following relation:
\be{eq:ZajfmanAlpha}
	2E_0 \alpha_p = \frac{1-\alpha}{1+\alpha}
\ee
It should be noted that $\frac{\partial T}{\partial E_0}$ yields
essentially the same information as $\frac{\partial T}{\partial p_0}$
about the sign of the trap's dispersion.

\section{Two non interacting ions}
\label{sec:TwoNonInteracting}

Before we look at the coupled motion of two interacting ions, we
will study the non interacting case, where we set $W \equiv 0$ in
\eq{eq:Hamiltonian}. Adding the interaction will then modify this
model and the relevant mechanism will become clearer.

Instead of solving the actual equations of motion for the two ions we
"take apart" the trap and follow the evolution of the relative
coordinate through the three regions of the trap (cf. figure
\ref{fig:TrapGeom}): (i) the central field free part, (ii) the mirrors
and (iii) the kink region, which connects these two. The evolution of
the relative coordinate in each of these parts can then be described
by two--by--two mappings of the relative coordinate and momentum. The
composite mapping, built up of these building blocks, describes the
relevant dynamics.

\subsection{Mapping the free motion}
\label{sec:MappFree} 

The trap potential $V(x)$ \eq{eq:trapPotential} is flat in the central
part of the trap and linear in the mirrors; in these regions the
Hamiltonian \eq{eq:Hamiltonian} can be separated into CM and relative
coordinates.

Without the ions' interaction, the relative momentum $p$ is constant
in both the central part and in the mirrors and the distance $x$
evolves freely as
\be{eq:FreeEvolution}
	x(t) = x(0) + \frac{p}{\mu}\: t. 
\ee
As the times spent in the mirror and in the central part, $T_m$ and
$T_f=\alpha T_m$, are fixed by the CM motion (see equations
(\ref{eq:TimeInTheMirror}) and (\ref{eq:TimeInTheFlat})), we define
two mappings $\mathcal{F}$ and $\mathcal{M}$ that describe this free
evolution of the relative coordinate in the central part and in the
mirrors, respectively. They differ only in their time duration:
%
\bea
	\mathcal{F}: 
	\left( \begin{array}{c} x \\ p \end{array} \right) & \mapsto &
	\left( \begin{array}{c} x' \\ p' \end{array} \right) = 
	\left( \begin{array}{c} x + \frac{\alpha T_m}{\mu}p \\ p \end{array} 
	\right)
	\label{eq:MappFree}
	\\
	\mathcal{M}: 
	\left( \begin{array}{c} x \\ p \end{array} \right) & \mapsto &
	\left( \begin{array}{c} x' \\ p' \end{array} \right) = 
	\left( \begin{array}{c} x + \frac{T_m}{\mu}p \\ p \end{array} \right)
	\label{eq:MappMirror}
\eea
Their interpretation is the following: if the ions have the relative 
coordinates $(x, p)$, when the CM enters, e.g., the mirror, these will 
have evolved to $(x', p') = \mathcal{M} (x,p)$, when the CM leaves this 
region again.

\subsection{The momentum kick approximation}
\label{sec:MomentumKick}

\epsfxsize=6cm
\epsfysize=3cm
\begin{figure}[t]
	\epsffile{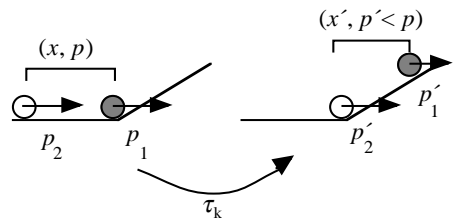}
	\caption{During the time $\tau_k$, when the two ions are in
	different parts of the trap potential, they experience a different
	acceleration due to the mirror potential. Integrating the
	difference over $\tau_k$ results in a compressing change of the
	relative momentum \eq{eq:MomentumKick}.}
	\label{fig:MomentumKick}
\end{figure}

For that part of the motion when the two ions are in different parts
of the trap potential, the CM and the relative motion are coupled for
a time $\tau_k$. This is the time between when the first ion crosses
the kink between the flat part and one of the mirrors at $\pm
\frac{L}{2}$ and when the second one passes that point (see figure
\ref{fig:MomentumKick}). During that time the ions are accelerated
relative to each other by the trap potential. For the case of incoming
ions, which is depicted in figure \ref{fig:MomentumKick}, this time is:
%
\[
	\tau_k = \frac{x}{p_2 / m}
\]
After this time the momentum $p_2$ will be the same, $p_2'=p_2$, but
$p_1'$ has become
%
\[
	p_1' = p_1 + \tau_k \left( -\frac{\partial V}{\partial x_1} \right) 
         = p_1 - \frac{x m}{p_2} F
\]
The two ions have nearly the same velocity, so $\frac{p_2}{m}$ is
approximated by $\frac{P}{M}$, the velocity of the CM.

If the energy of the two ions of about 4.2~keV differs by 2~eV in the
laboratory frame, their momenta will differ by about 1~a.u.: replacing
$\frac{p_2}{m}$ by $\frac{P}{M}$ then introduces an error on the order
of $\mathcal{O}(10^{-5})$.

The relative momentum $p = \frac{p_1 - p_2}{2}$ now changes as (cf.
\eq{eq:TimeInTheMirror})
\be{eq:MomentumKick}
	p' = p - \frac{F M}{2 P}x = p - \frac{2 \mu}{T_m}x
\ee
This formula is valid for outgoing ions, too. Through the special 
geometry of the trap the momentum transfer and the time spent in the 
mirror are intimately related to each other.

The relative distance $x$ changes, too, but assuming an initial
distance of, e.g., $x = 10^7$ a.u., only by a factor of $1 -
\mathcal{O}(10^{-3})$; this change of $x$ will be neglected. We also
assume that the CM momentum is not affected by this momentum transfer,
which is true to the order of $\mathcal{O}(10^{-3})$. 

As the ions --- in our approximation --- do not move relative to each
other during $\tau_k$, the motion through the kink region has the
effect of an instantaneous momentum "kick" in the relative coordinate,
induced by the coupling of the CM and the relative coordinate. This
momentum "kick" always pushes the two ions together. We consequently
describe the kink by the following mapping $\mathcal{K}$:
\be{eq:MappKick}
	\mathcal{K}: 
	\left( \begin{array}{c} x \\ p \end{array} \right) \mapsto
	\left( \begin{array}{c} x' \\ p' \end{array} \right) = 
	\left( \begin{array}{c} x  \\ p - \frac{2 \mu}{T_m}x \end{array} 
	\right)
\ee

\subsection{The composite mapping --- the dispersion of the trap}

\begin{figure}[t]
	\epsffile{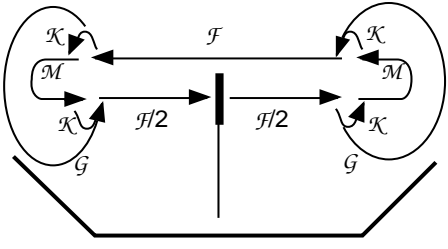}
	\caption{Sketch of the sequence of mappings during one period of
	the motion of two non interacting ions \eq{eq:NonInterMapp}. The
	trap is symbolized by the potential and the stylized pickup.}
	\label{fig:NonInterMapp}
\end{figure}

With the mappings $\mathcal{F}$, $\mathcal{M}$ and $\mathcal{K}$ in
hand we are now able to describe the motion of two non interacting
ions in the trap. One complete period consists of a sequence of these
mappings, cf.~figure \ref{fig:NonInterMapp}. We start the cycle in the
middle of the central region, where in the experiment the pickup is
located. The composite map, denoted by $\mathcal{P}_{\non}$, then
describes how the observed distance between the two ions changes
between successive measurements. It has the following form, with
$\mathcal{F}_{\frac{1}{2}}$ describing the free motion for a time
$\frac{T_f}{2}$ (cf. \eq{eq:MappFree}):
%
\bea
	\!\!\!\mathcal{P}_{\non} & \!\! = \!\!& \mathcal{F}_{\frac{1}{2}} 
		\!\otimes 
		\underbrace{\mathcal{K} \otimes \mathcal{M} \otimes \mathcal{K}}
		\otimes \mathcal{F} \!\otimes 
		\underbrace{\mathcal{K} \otimes \mathcal{M} \otimes \mathcal{K}}
		\otimes \mathcal{F}_{\frac{1}{2}} \\
		\label{eq:NonInterMapp}
	&\!\! = \!\!& \mathcal{F}_{\frac{1}{2}} \!\otimes 
		\hspace{0.85cm} \mathcal{G} \hspace{0.7cm}
		\otimes \mathcal{F} \!\otimes 
		\hspace{0.85cm} \mathcal{G} \hspace{0.65cm}
		\otimes \mathcal{F}_{\frac{1}{2}}
		\label{eq:NonInterMappG}
\eea
%
\epsfxsize=6cm
\epsfysize=3.375cm
In the second line the effects of the mirror and the two adjacent 
momentum kicks are combined into the mapping $\mathcal{G}$. We will 
later see that the main synchronization process can be understood from 
this part alone.
The composite mapping $\mathcal{P}_{\non}$ finally simplifies to
\be{eq:MappNonInter}
	\mathcal{P}_{\non}: 
	\left( \begin{array}{c} x \\ p \end{array} \right) \mapsto
	\left( \begin{array}{c} x' \\ p' \end{array} \right) = 
	\left( \begin{array}{c} x - \frac{2 T_m}{\mu} (1- \alpha) p \\ p 
		   \end{array} \right)
\ee
From energy conservation we know that the momenta $p_1$ and $p_2$, and
consequently $p$, have to be the same, when each of the independent
ions has completed the period. The increase in distance, already
calculated from the dispersion $\frac{\partial T}{\partial p_0}$
\eq{eq:Dispersion} is also reproduced by $\mathcal{P}_{\non}$.

Our mapping model therefore describes the evolution of the relative
distance between two independent ions with the same accuracy of
$\mathcal{O}(10^{-5})$ as the linearized dispersion
\eq{eq:Dispersion}.

\subsection{The mirrors' map: 
$\mathcal{G} = \mathcal{K}\otimes\mathcal{M}\otimes\mathcal{K}$}
\label{sec:MirrorMap}

\begin{figure}[t]
	\epsffile{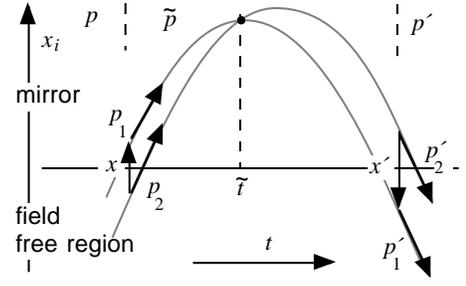}
	\caption{Sketch of the trajectories of two non interacting ions on
	their way from the central field free region of the trap through
	the mirror back into the central part. The horizontal axis is
	time, the vertical axis denotes the position of the ions in the
	trap. Ion 2 is here the faster one. The initial relative momentum
	is $p$, after the first kick it is denoted by $\tilde{p}$ and
	after the second kick by $p'$. The time when the ions'
	trajectories cross is denoted by $\tilde{t}$.}
	\label{fig:MomentaFlip}
\end{figure}

Now we will look at the above defined mapping $\mathcal{G}$
\eq{eq:NonInterMappG}. From energy conservation and geometrical
considerations we see that --- without the ions' interaction --- the
total effect of the mirror is to turn around the direction of the two
momenta $p_1$ and $p_2$, and therefore of $p$ (see figure
\ref{fig:MomentaFlip}):
%
\epsfxsize=6cm
\epsfysize=3.75cm
\be{eq:MomentaFlip}
	\left.
	\begin{array}{c} p_1' = -p_1 \\ p_2' = -p_2 \end{array}
	\right\} \Rightarrow \quad 
	p' = \frac{p_1' - p_2'}{2} = -p
\ee
The relative momentum changes by $\Delta p = p'-p = -2p$. Each of the 
two momentum kicks is proportional to the corresponding distance 
\eq{eq:MomentumKick}, $x$ at the entrance and $x'$ at the exit of the 
mirror; the intermediate and final momenta, $\tilde{p}$ 
and $p'$, are therefore
\be{eq:TwoKicks}
	\tilde{p}=p-\frac{2\mu}{T_m}x \quad \mbox{and} \quad
	p'=\tilde{p}-\frac{2\mu}{T_m}x' \quad
\ee
From the above equations \eq{eq:MomentaFlip} and \eq{eq:TwoKicks} we
derive a relation between the distances $x$ and $x'$ and the momentum
$p = -p'$:
%
\[
	x + x' = \frac{T_m}{\mu} p
\]
This relation can also be obtained by evaluating $\mathcal{G} =
\mathcal{K}\otimes\mathcal{M}\otimes\mathcal{K}$ directly:
%
\begin{equation}
	\mathcal{G}: 
	\left( \begin{array}{c} x \\ p \end{array} \right) \mapsto
	\left( \begin{array}{c} x' \\ p' \end{array} \right) = 
	\left( \begin{array}{c} - x + \frac{T_m}{\mu} p \\ -p 
		   \end{array} \right)
\end{equation}

It should be noted that with the experimental parameters of
\tbl{tbl:Dimensions}, the change of $x$ in the mirror is about a
factor of 10 bigger than $\Delta x$ after one complete period
\eq{eq:IncreaseDistDisp}. Most of the mirror's effect, however, is
compensated for by the motion in the central part. At $\frac{\partial
T}{\partial p_0}=0$, which corresponds to $\alpha=1$, both
contributions exactly cancel each other.

\epsfxsize=17.2cm
\epsfysize=6.5cm
\begin{figure*}[t]
	\epsffile{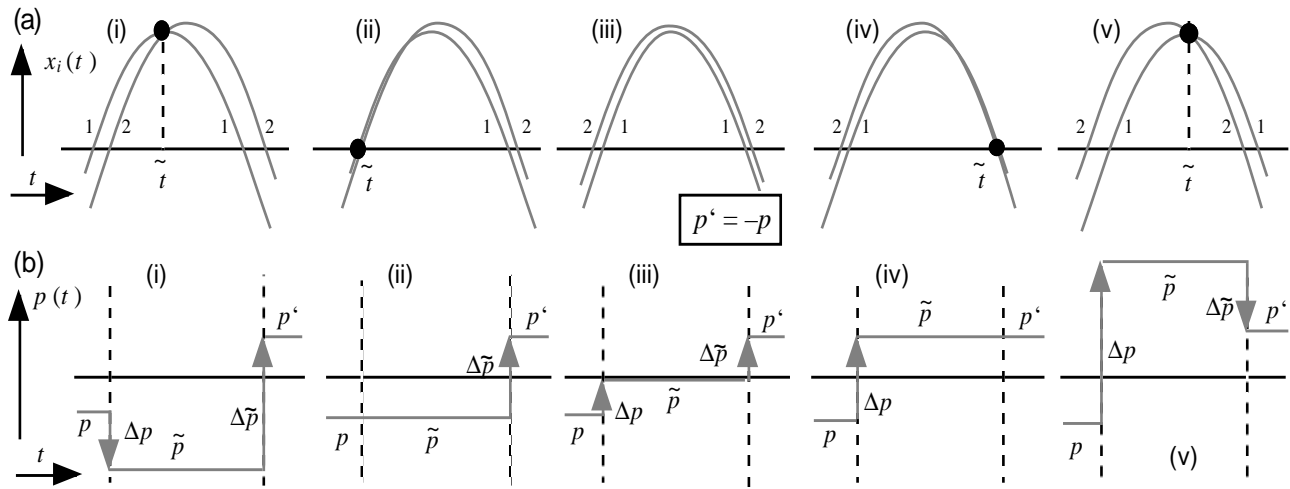}
	\caption{The different types of trajectories that occur in the
	mirror region for different initial values; for an explanation
	please see text.}
	\label{fig:TrajTypes}
\end{figure*}
In figure \ref{fig:MomentaFlip} the time $\tilde{t}$ denotes the
crossing of the two ions' trajectories. It is defined by
%
\[
	\tilde{x}(\tilde{t}) = x + \frac{\tilde{p}}{\mu}\; \tilde{t} = 0
\]
and evaluates to
\be{eq:NonInterCrossT}
	\tilde{t} = \tilde{t}(x, p) = \frac{T_m \mu x}{2\mu x - T_m p}
\ee
The limits for $x \to \pm \infty$ or $p=0$ are
$\tilde{t}=\frac{T_m}{2}$. For finite $p$ and $\frac{x}{p} < 0$, $0
\leq \tilde{t} \leq \frac{T_m}{2}$; for $\frac{x}{p} \geq
\frac{T_m}{\mu}$ we have $\frac{T_m}{2} \leq \tilde{t} \leq T_m$. In
the intermediate region $\tilde{t}$ is either negative or bigger than
$T_m$, with a singularity at $\frac{T_m p}{2 \mu}$. In that case there
is no collision inside the mirror, as \eq{eq:NonInterCrossT} is only
defined in the mirror region.

Equation \eq{eq:NonInterCrossT} is meaningful only as long as both
ions spend time together in the mirror: if the first ion has already
left the mirror when the second one arrives, the above treatment is
not valid. This happens when the ions are further away from each other
than
\be{eq:XMaxCross}
	x_{\mbox{\scriptsize max}} = T_m \frac{P}{M} = \frac{L}{\alpha}
\ee
Without interaction the two ions' trajectories cross twice during a
given period. This happens either in the mirror or in the central
region. The probability $w_F$ for the latter is given by the ratio
between the relative distance for which there is no solution of
\eq{eq:NonInterCrossT} inside the mirror, $\frac{T_m p}{\mu}$, and the
distance $x_{\mbox{\scriptsize max}}$ \eq{eq:XMaxCross}, for which
equation \eq{eq:NonInterCrossT} is defined. With \eq{eq:TimeInTheFlat}
it evaluates to
\be{eq:FlatProb}
	w_F = \frac{Mp}{\mu P} = \frac{4p}{P}
\ee
As $p \ll P$, a collision of the two ions in the central part is a
highly unlikely event. With the parameters given, $w_F$ is on the order
of only $\mathcal{O}(10^{-3})$.

The different possibilities can be visualized by shifting the two
(parabolic) trajectories of the ions in figure \ref{fig:MomentaFlip}
against each other: as they are plotted in figure
\ref{fig:MomentaFlip} we have $0 < \tilde{t} < \frac{T_m}{2}$. This is
case (i) of figure \ref{fig:TrajTypes}. If now the trajectory of ion 1
is moved to the right, the crossing time $\tilde{t}$ will slide to the
left, until both trajectories intersect at their entrance into the
mirror, which means $\tilde{t} = 0$, depicted in case (ii) of figure
\ref{fig:TrajTypes}a. When trajectory 1 is shifted further to the
right, case (iii), the ions do not cross inside the mirror (cf. the
explanation above and equation \eq{eq:FlatProb}). The crossing point
reappears at $\tilde{t}=T_m$ (iv), i.e., when the ions leave the
mirror, and from there it proceeds back in the direction of
$\frac{T_m}{2}$ (v). These processes are the framework into which we
later incorporate the ions' interaction.

Fig.~\ref{fig:TrajTypes}(b) plots the relative momentum $p(t)$ during
the course of $\mathcal{G}$. In all cases the final momentum is
$p'=-p$. For cases (i) and (ii) the momentum changes sign only with
the second kick, while for cases (iv) and (v) it is reversed already
with the first kick. When the crossing takes with $\tilde{t}<
\frac{T_m}{2}$ the second kick is the larger one, while if $\tilde{t}
> \frac{T_m}{2}$, the first kick is stronger. In all these cases the
two kicks have opposite direction.

In the above cases $x(t)$ changes sign in the mirror at $t=\tilde{t}$. In
the special case (iii) $x$ does not change its sign and both kicks
work in the same direction; here the momentum in the mirror is less
than before and after it. In this regime the faster ion enters the
mirror first and leaves it second.

It should be emphasized that, without interaction, the relative
momentum in the central part of the trap is the same after each
period. This is the important conserved quantity. The relative
momentum in the mirror, on the other hand, is a measure of the time
delay or phase lag of the two ions on their orbit.

\section{Adding the ion--ion interaction}
\label{sec:AddingInteraction}

We now incorporate the (repulsive) interaction between the two ions
into our model. We will see that the repulsive interaction can be
modelled as a time delay and we will then introduce this delay into
the mappings.

\subsection{The time delay}

When both ions are in one of the mirrors or in the central part, the
relative coordinate is decoupled from the CM motion (see section
\ref{sec:MappFree}). The free motion in the relative coordinate is now
modified by the interaction $W(x)$ \eq{eq:FreeEvolution}. Any
collision of the ions is elastic due to the conservative Coulomb
interaction. As we are only interested in the final values of $x$ and
$p$, but not in the actual solution of the equations of motion, we
adopt a central idea of scattering theory: the whole effect of the
ions' collision is described by a phase shift, or, for our treatment,
a time delay $\tau$ \cite{TAYLOR}. This delay modifies the propagation
time for the freely evolving relative coordinate:
\be{eq:FreeDelayed}
	x(T) = x(0) + \frac{p}{\mu} (T - \tau)
\ee
In this ansatz $T$ stands for either $T_m$ or $T_f$. The time delay
$\tau$ is clearly a function of the interaction potential and the
initial distance and momentum. It can be positive or negative; if the
relative motion is delayed we have $\tau > 0$.

\subsection{Modifying the mappings}

With the interaction added we now use \eq{eq:FreeDelayed} instead of
\eq{eq:FreeEvolution} to map the relative coordinate in the mirrors or
the trap's central part, respectively.

If the initial and final distances are not in the asymptotic region,
where $W(x) = 0$, we have to adjust $p'$, too, to conserve the energy
of relative motion. But as our treatment does not depend on a special
form of $W$, we will for the following assume that the collision
between the two ions takes much less time than $T_m$ or $T_f$ and that
before and after the collision the interaction between the ions can be
neglected.

A time delay $\tau_m$ in the mirror region may be incorporated directly
into \eq{eq:MappMirror}. It modifies the relative coordinate:
\be{eq:MappMirrorMod}
	\mathcal{M}': 
	\left( \begin{array}{c} x \\ p \end{array} \right) \mapsto
	\left( \begin{array}{c} x' \\ p' \end{array} \right) = 
	\left( \begin{array}{c} x + \frac{T_m-\tau_m}{\mu}p \\ p \end{array} \right)
\ee
The mapping now is a functional of the still unspecified delay $\tau_m 
= \tau_m(W; x, p)$, which itself depends on the actual form of the 
interaction potential. 

In the non interacting case the mapping in the central part is
described by \eq{eq:MappFree}. Recall that the ratio of $T_f$ and
$T_m$ defines the parameter $\alpha$, which in turn is related to the
dispersion of the trap \eq{eq:Dispersion}. It will be convenient,
therefore, to incorporate a time delay $\tau_f$ into a modified
$\alpha'$, defined by:
%
\begin{equation}
	\alpha T_m - \tau_f = (\alpha - \frac{\tau_f}{T_m}) T_m
	=: \alpha' T_m 
\end{equation}
The mapping $\mathcal{F}$ is consequently modified to:
\be{eq:MappFreeMod}
	\mathcal{F}': 
	\left( \begin{array}{c} x \\ p \end{array} \right) \mapsto
	\left( \begin{array}{c} x' \\ p' \end{array} \right) = 
	\left( \begin{array}{c} x + \frac{\alpha' T_m}{\mu}p \\ p \end{array} \right)
\ee

During the short interval of the momentum kick we assume the ions do
not move relative to each other (see section \ref{sec:MomentumKick}).
Also the trap potential is much stronger than the weak ion--ion
interaction. Consequently the mapping $\mathcal{K}$ \eq{eq:MappKick}
remains unchanged.

With these modifications the composite mapping for the whole period,
starting and ending at the pickup in the center of the trap, becomes a
functional of the delays $\tau_m$ and $\tau_f$ (cf.
\eq{eq:NonInterMapp}):
\be{eq:InterMapp}
	\mathcal{P}' =
		\!\!\mathcal{F}_{\frac{1}{2}}' \!\otimes 
		\mathcal{K}\! \otimes \mathcal{M}' \otimes \mathcal{K}
		\!\otimes \mathcal{F}' \!\otimes
		\mathcal{K}\! \otimes \mathcal{M}' \otimes \mathcal{K}
		\!\otimes \mathcal{F}_{\frac{1}{2}}'
\ee
Here $\mathcal{F}_{\frac{1}{2}}'$ denotes the propagation through only
half of the central part, i.e., between the pickup and one of the
mirrors: $\mathcal{F}' = \mathcal{F}_{\frac{1}{2}}' \otimes
\mathcal{F}_{\frac{1}{2}}'$.

\section{Linear stability analysis}
\label{sec:LinearStability} 

Now that we have set up all parts of the model we will quantify the
stability of the composite mapping $\mathcal{P}'$ \eq{eq:InterMapp}
and determine the conditions under which synchronization of the two
ions' motion occurs.

We follow the usual lines: the mappings are linearized around the
initial values $(x, p)$ by calculating the Jacobi matrix of partial
derivatives:
%
\[ 
	J = J(\tau_m, \tau_f) = 
	\left( \begin{array}{cc}
		\frac{\partial x'}{\partial x} &\;\; \frac{\partial p'}{\partial x}  
			\\[0.2cm]
		\frac{\partial x'}{\partial p} &\;\; \frac{\partial p'}{\partial p}
	\end{array} \right) .
\]
If the eigenvalues $\lambda$ of $J$ are complex with a length of 1,
then the mapping is stable and the ions' motion synchronized; if the
$\lambda$ are real it is unstable and the motion diffusive
\cite{PER82}. According to equations (\ref{eq:MappKick}),
(\ref{eq:MappMirrorMod}) and (\ref{eq:MappFreeMod}) the eigenvalues
are functions of the trap parameters $T_m$, $\alpha$ and $\mu$ and of
the time delays $\tau_m$ and $\tau_f$.

\subsection{The elementary maps}

Through the linearization the time delays become constants. With the
constant $\tau_m$ and $\tau_f$, and, consequently, the constant
$\alpha'$, all the maps involved are linear and their Jacobians can be
represented by $2 \times 2$ matrices. They have the following forms:
\bea
	\mathcal{F}' \Rightarrow \mathbf{F} & = \left[
		\begin{array}{cc} 1 &\;\; \frac{\alpha' T_m}{\mu} \\[0.1cm] 0 & 1
		\end{array} \right] &\;\; \\
	\mathcal{M}' \Rightarrow \mathbf{M} & = \left[
		\begin{array}{cc} 1 &\;\; \frac{T_m-\tau_m}{\mu} \\[0.1cm] 0 & 1
		\end{array} \right] &\;\; \\
	\mathcal{K} \Rightarrow \mathbf{K} & = \left[
		\begin{array}{cc} 1 &\;\; 0 \\[0.1cm] -\frac{2\mu}{T_m} & 1
		\end{array} \right] &\;\; 
\eea
The determinant of each of these matrices is equal to 1. Moreover, they all have
a double eigenvalue of 1 and a fixpoint at $(x,p)=0$, independent of
the values of $T_m$, $\mu$, $\alpha'$ and $\tau_m$.

The composite mappings are represented by the corresponding products
of the elementary matrices: $\mathbf{G} = \mathbf{KMK}$ and
$\mathbf{P} =
\mathbf{F}_{\frac{1}{2}}\mathbf{KMKFKMKF}_{\frac{1}{2}}$, where
$\mathbf{F}_{\frac{1}{2}}$ corresponds to $\mathcal{F}_{\frac{1}{2}}'$
(cf. \eq{eq:InterMapp}). Since $\mathbf{F}$, $\mathbf{M}$ and
$\mathbf{K}$ all have a fixed point at $(x,p)=0$, it follows that
$\mathbf{G}$ and $\mathbf{F}$ must as well.

\subsection{Stability of the complete mapping $\mathcal{P}$}

The Jacobian for the complete period can be grouped into the modified
mirror $\mathbf{G}= \mathbf{KMK}$ and the field free central region
$\mathbf{F}$:
$\mathbf{P}=\mathbf{F}_{\frac{1}{2}}\mathbf{GFGF}_{\frac{1}{2}}$. The
stability criterion can, due to this repetitive structure, already be
inferred from $\mathbf{P}_{\frac{1}{2}} =
\mathbf{F}_{\frac{1}{2}}\mathbf{GF}_{\frac{1}{2}}$. Its explicit form
is, 
\be{eq:HalfP}
	\mathbf{P}_{\frac{1}{2}} = \left[ \begin{array}{cc}
		-1 + 2\epsilon(1-\alpha) & \;
			\frac{T_m}{\mu}(1-\alpha)(1-\epsilon +\alpha\epsilon) \\[0.2cm]
		-\frac{4\mu \epsilon}{T_m} & -1 + 2\epsilon(1-\alpha)
	\end{array} \right]
\ee
where we have defined $\epsilon = \frac{\tau_m}{T_m}$.

The determinant of $\mathbf{P}_{\frac{1}{2}}$ is 1. In section
\ref{sec:MirrorMap} we showed that the ions collide \emph{either} in
the mirror \emph{or} in the central part; therefore one of the delays
vanishes. These two cases correspond to either $\alpha'=\alpha$ or
$\tau_m=0$, respectively. For $\alpha'=\alpha$ \emph{and} $\tau_m=0$ the non
interacting case is recovered: $\mathbf{P} = \mathbf{P}_{\frac{1}{2}}
\mathbf{P}_{\frac{1}{2}} = \mathbf{P}_{\non}$, where $\mathbf{P}_{\non}$
is the matrix corresponding to the mapping of equation
\eq{eq:MappNonInter}.

\subsubsection{Collisions in the central part}

First we will consider the case of $\tau_m = 0$, i.e., the collisions
take place in the central part. As discussed above this case is rare.
Then $\mathbf{P}=\mathbf{P}_{\frac{1}{2}} \times
\mathbf{P}_{\frac{1}{2}}$ reduces to the dispersion of the non
interacting case \eq{eq:MappNonInter}, but with the externally defined
$\alpha$ replaced by the modified $\alpha' = \alpha -
\frac{\tau_f}{T_m}$. Consequently $\mathbf{P}_{\frac{1}{2}}$ has a
double eigenvalue of $-1$, i.e., the distance between the ions grows
linearly in time according to equation \eq{eq:IncreaseDistDisp} with
$\alpha$ replaced by $\alpha'$. Which ion is faster is determined by
whether $\alpha'$ is bigger or smaller than 1.

As in the non interacting case, the eigenvalues of the composite map
for the whole trap are $\lambda_{1/2}=1$: the ions separate linearly
in time, if the modified dispersion is not exactly $\alpha' = 1$.

For $\alpha < 1$, which is the experimentally observed condition for
bunching, a negative delay $\tau_f$ may bring the trap into the regime
where the "effective" dispersion $\alpha'$ becomes 1 and both ions
have the same period. Collisions in the central part may therefore
synchronize the ions by modifying the dispersion of the trap: their
differing momenta remain unchanged, but the distance is, due to the
vanishing dispersion, the same every time they pass the trap's center.
The possibility of this type of synchronization is small: first
collisions in the central part are rare events (see equation
\eq{eq:FlatProb}) and second, with the eigenvalue of 1 the
synchronization, if it occurs, is not stable against perturbations.

\subsubsection{Collisions in the mirror}
\label{sec:MirrorCollisions} 

In the other, much more frequent case, where the ions collide in the
mirrors, i.e., $\alpha' = \alpha$ and $\tau_m \neq 0$, the eigenvalues
have the form
\be{eq:LambdaP12Raw}
	\lambda_{1/2} = -1 + 2\epsilon(1-\alpha)
		\pm 2 \sqrt{\epsilon^2 (1-\alpha)^2 - 
			\epsilon(1-\alpha)}
		\,.
\ee
The unmodified $\alpha$ describes the trap's original dispersion. It
is now convenient to define another parameter
\be{eq:GammaDef}
	\gamma = \epsilon(1-\alpha). 
\ee
Then the eigenvalues \eq{eq:LambdaP12Raw} simplify to
\be{eq:LambdaP12}
	\lambda_{1/2} = -1 + 2\gamma \pm 2 \sqrt{\gamma^2 - \gamma},
\ee
The argument of the square root, $\gamma^2 - \gamma$, is negative for
$0< \gamma < 1$. In that region the eigenvalues are consequently
complex with a length of 1 and they are real for $\gamma < 0$.

The important parameter, as we see, is composed of the time delay
$\tau_m$ in the mirror and the dispersion of the trap, described by
$\alpha$ \eq{eq:Dispersion}. The sign of this parameter $\gamma$
determines if the relative motion of the two ions is bounded or not:
for $\gamma > 0$, which requires either $\tau_m>0$ and $\alpha<1$ or
$\tau_m<0$ and $\alpha>1$, the eigenvalues $\lambda$ are complex and
the ions' motion is synchronized. For $\gamma < 0$ the eigenvalues are
real and the ions separate faster than without interaction. This
happens for $\tau_m>0$ and $\alpha>1$ or for $\tau_m<0$ and
$\alpha<1$.

\begin{table}[tbp]
	\centering
	\begin{tabular}{|c||c|c|}
		\hline
		 & $\tau_m>0  $ & $\tau_m<0 $ \\
		\hline
		\hline
		$\ba{c} T_m>T_f \\ \Rightarrow \alpha<1 \ea $ & 
			$\ba{c} \gamma>0 \\ \Rightarrow \lambda \mbox{ complex} \\ 
			\mbox{synchronization} \ea $& 
			$\ba{c} \gamma<0 \\ \Rightarrow \lambda \mbox{ real} \\\mbox{ 
			diffusion} \ea $\\
		\hline
		$\ba{c} T_m<T_f \\ \Rightarrow \alpha>1 \ea $ & 
			$\ba{c} \gamma<0 \\ \Rightarrow \lambda \mbox{ real} \\\mbox{ 
			diffusion} \ea $ &
			$\ba{c} \gamma>0 \\ \Rightarrow \lambda \mbox{ complex} \\ 
			\mbox{synchronization} \ea $ \\
		\hline
	\end{tabular}
	\caption{Overview of the eigenvalues $\lambda$ of the composite mapping 
	matrix $\mathbf{P}_\frac{1}{2}$
	\eq{eq:HalfP} for collisions in the mirror and the corresponding
	behavior of the ions. For $\gamma = 0$, i.e., $\tau_m=0$ or
	$\alpha=1$, which corresponds to either collisions in the central
	part or no interaction at all, the eigenvalues are $-1$ and the
	ions' separation in phase space grows linearly in time.}
	\label{tbl:EigenValues}
\end{table}

For very small $\gamma$, i.e., $\tau_m \ll T_m$ or $\alpha \approx 1$,
we can neglect $2\gamma$ compared to 1 in \eq{eq:LambdaP12} and
approximate the square root by $\sqrt{\gamma^2 - \gamma} \approx
\sqrt{-\gamma}$. The eigenvalues are then of the form
%
\begin{equation}
	\lim_{\gamma \to 0} \lambda_{1/2} = -1 \pm \sqrt{-\gamma}
\end{equation}
Consequently the transition between real and complex eigenvalues is
very abrupt and the stability changes completely as the delay or the
dispersion changes its sign. Table \ref{tbl:EigenValues} gives an
overview of the behavior of the eigenvalues for the different values
that $\alpha$ and $\tau_m$ can take for collisions in the mirror.

The stability analysis is consistent with the experimental observation
and the findings of reference \cite{STR02} that synchronization occurs
in the region of $\frac{\partial T}{\partial p_0} > 0$, which in our
parametrization corresponds to $\alpha < 1$, or according to
\cite{PED02b} to $2E \alpha_p > 0$ \eq{eq:ZajfmanAlpha}: to achieve
bunching, ions with higher energy must have a longer period. But this
criterion alone is not enough; it has to be accompanied by a time
delay in the mirror, which in turn requires a repulsive ion--ion
interaction.

For the other scenario, where the mapping is stable for $\alpha > 1$,
the time delay has to be negative. This is possible if the ions have
an attractive potential, e.g., if their charges have opposite signs,
but then the electrostatic field of the mirrors could keep only one of
the ions trapped; their motion would not even have a chance to
synchronize. For two or more equal ions only a repulsive Coulomb
interaction is possible; any other interaction like the attractive van
der Waals interaction is much weaker and can not compete with the
Coulomb repulsion. A negative time delay is also possible when the
ions' repulsion becomes comparable to the guiding mirror field: then a
repulsion between the ions can lead to a negative delay --- the ions
``bounce off'' each other. But in this case the ions' repulsion would
dominate the dynamics and it is questionable if the very weak trap
potential would be able to actually trap the ions.

\section{Explaining the synchronization mechanism}
\label{sec:Explaining} 

From the stability analysis we identify two different mechanisms that
can synchronize the motion of two ions: when the ions collide in the
mirror, the relation between the momentum kicks is unbalanced, and
when collisions take place in the central part, they modify the
effective dispersion. We will now have a closer look at the
microscopic dynamics in these two regimes.

\subsection{Collisions in the mirror}
\label{sec:ExplainMirror} 

\epsfxsize=8.6cm
\epsfysize=7cm
\begin{figure}[t]
	\epsffile{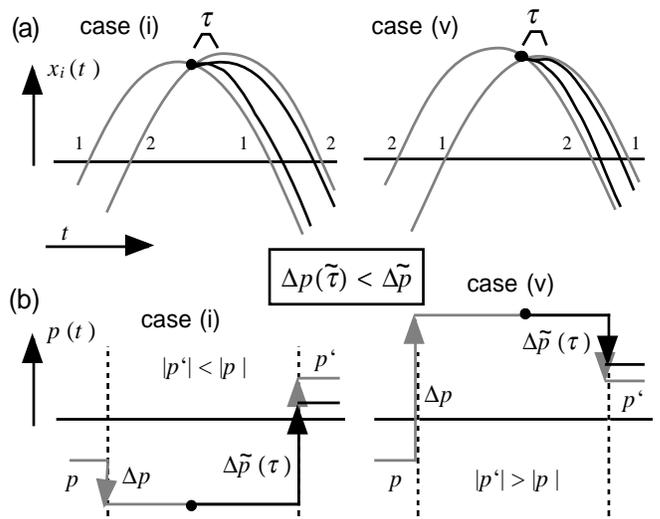}
	\caption{(a) The ions' positions in the mirror vs. time for case
	(i), where $\tilde{t}<T_m/2$ and for case (v), where
	$\tilde{t}>T_m/2$ (cf. figure \ref{fig:TrajTypes}). (b) The
	corresponding evolution of the relative momentum. A time delay
	$\tau_m$ in the mirror leaves the ions less time to separate after
	their collision. Consequently the second momentum kick is smaller
	than without interaction (gray lines).}
	\label{fig:ModTrajTypes}
\end{figure}

To explain how the collisions in the mirror synchronize the ions, it
is sufficient to look at the modified mirror map $\mathcal{G}' =
\mathcal{K}\otimes \mathcal{M}'\otimes \mathcal{K}$. Without the
interaction, $\mathcal{G}$ exactly turns around the relative momentum
of the two independent ions: the two momentum kicks add up to $-2p$.
During the interval in the mirror the ions approach and reseparate.
When the repulsive interaction is added the ions experience a time
delay and, consequently, have less time to separate: the second
momentum kick \eq{eq:MomentumKick} will be smaller than without
interaction. This is depicted in figure \ref{fig:ModTrajTypes}. The
two sets of trajectories correspond to the configurations (i) and (v)
of figure \ref{fig:TrajTypes}, which shows the non interacting case.
Now the two kicks, which both are proportional to the corresponding
relative distance, will not add up to $-2p$ any more and the relative
momentum after the second kick will be either smaller (figure
\ref{fig:ModTrajTypes}, case (i)) or bigger (case (v)) than $-p$, depending on
whether the slower or the faster ion enters the mirror first. In
either case momentum is effectively transferred from the second onto
the first ion, due to the modified coupling between the relative and
the CM motion.

If the first ion is the faster one it will gain additional energy
through this momentum transfer. Because the period increases with
energy, the distance between the two ions will shrink until the slower
one overtakes the faster one. Then momentum is transferred from the
faster ion, which is now the second one, onto the slower first one,
until the first ion again will be the one with the higher momentum.
This process repeats itself over and over, with the ions effectively
orbiting around each other in their relative coordinate, stably
interlocking and synchronizing their motion.

In the non interacting case the momentum kicks and the time spent in
the mirror are intimately related (see \eq{eq:NonInterCrossT}). This
balance is now offset by the ion--ion interaction. In this way the very
small repulsion is amplified to counterbalance the trap's dispersion
and suppress the separation of the ions.

\subsection{Collisions in the central part}
\label{sec:ExplainCentral} 

When the ions collide in the central part of the trap the mechanism is
completely different. As shown in section \ref{sec:MirrorMap}, in this
regime the faster ion enters the mirror first and leaves second 
(see figure \ref{fig:TrajTypes}(iii)). For this mode to be stationary
the dynamics has to be symmetric in both mirrors; consequently the
ions have to exchange their roles in the collision in the central part
of the trap: the first slower ion is accelerated by the faster
second, which comes from behind. With this mechanism, which was dubbed
``bricking motion'' by Zajfman \etal \cite{DANIEL}, each of the ions
is the faster for one half period and the slower one on the other side
of the trap. This way both ions have the same energy, and therefore
the same period on average, even for $\alpha \neq 1$.

\section{Illustrations and extensions}
\label{sec:Numericals} 

In the previous sections we have derived the conditions for stability
and explained the underlying mechanism of our mapping model. Now we
will check the validity of our description with regard to the two
central simplifications --- the instantaneous ion--ion interaction and
the constant slope of the mirror potentials.

\subsection{Poincar\'e maps for constant $\tau_m$}

In section \ref{sec:ExplainMirror} we explained that in the bunching
mode the ions oscillate around each other. This "orbiting" of the ions
can be conveniently depicted in a Poincar\'e section of the repeated
mapping $\mathcal{P}'$ \eq{eq:InterMapp} by plotting the relative
coordinate and momentum after each iteration.

\epsfxsize=8cm
\epsfysize=4.5cm
\begin{figure}[t]
	\epsffile{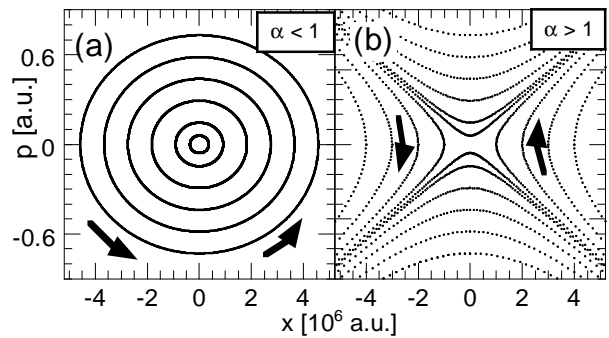}
	\caption{Poincar\'e section for the composite mapping
	$\mathcal{P}$ for a constant $\tau_m = 3\times 10^8$ a.u. Panel
	(a) shows the bunching and (b) the diffusive behavior. The arrows
	denote the direction of the phase space flow.}
	\label{fig:PoincConstTau}
\end{figure}
In section \ref{sec:LinearStability} we linearized $\mathcal{P}'$,
which resulted in a constant $\tau_m$. Figure \ref{fig:PoincConstTau}
shows the results of the repeated mapping for an arbitrarily chosen
$\tau_m = 3 \times 10^8$ a.u.: Panel (a) is calculated for $\alpha =
0.956$ based on the experimental values given in \tbl{tbl:Dimensions},
resulting in ellipses.

If we increase $\alpha$ to $1.046 = 1/ 0.956$ into the unstable
regime, with all the other parameters unchanged, the Poincar\'e
section becomes hyperbolic, see panel (b): the ions separate with a
speed that increases with their distance.

Note that the phase space flow in figure \ref{fig:PoincConstTau}(a) 
is in the direction opposite to that of an isolated particle 
in an external harmonic potential.

For a constant $\tau_m$ and $\alpha<1$, using the linearized mapping
$\mathbf{P}_{\frac{1}{2}}$ \eq{eq:HalfP} we can derive a conversion
factor between the maximal distance $x_{max}$ and the maximal momentum
$p_{max} = \frac{x_{max}}{\beta}$ of a Poincar\'e ellipse. It
evaluates to
\bea
	\beta & = & \frac{T_m}{2\mu}
	\sqrt{\frac{1-\alpha}{\epsilon} - (1-\alpha)^2}
		\label{eq:XPConversion} \\
	& = & \frac{T_m}{2\mu\epsilon}\sqrt{\gamma - \gamma^2}
\eea
%
The parameter $\gamma$ \eq{eq:GammaDef} is the same as 
defined in the context of the stability of the bunching mode (cf. section
\ref{sec:MirrorCollisions}), but the sign of the argument of the
square root is reversed. For the Poincar\'e sections shown above with
$\tau_m=3\times 10^8$ a.u. we get $\beta = 7.6\times 10^6$ a.u.

From equation \eq{eq:XPConversion} we see that $\beta$ decreases when
the trap is brought closer to the transition at $\alpha=1$:  the
maximal momentum increases until the point that the ellipses ``break
apart'' and become hyperbolas.

\subsection{Trajectory calculations: non instantaneous delay}

With the simple mapping model we were able to derive the condition for
a stable bunch and to give a microscopic explanation for the coupled
motion of the two ions. In order to verify that the discrete
description of the relative coordinate implied by the mapping is
accurate, we compare our results to trajectory calculations, i.e., to
the exact solution of the equations of motion of the two ions from the
full Hamiltonian \eq{eq:Hamiltonian}. 

To allow the ions to pass by each other in our one dimensional model
we replace the Coulomb repulsion $\frac{1}{|x|}$ between the ions by a
so called "softcore Coulomb interaction" \cite{Softcore}:
%
\[
	W(x) = \frac{1}{\sqrt{x^2 + d^2}}
\]
The "softcore parameter" or "impact parameter" $d$ is a measure of
how close the two ions have to come when passing each other; it
describes the diameter of the ion beam in the trap potential. For the
trajectory calculation we set this value to $d=10$ $\mu$m. This value
is much smaller than the real beam, but still the ions'
interaction is weaker than the trap potential by about four orders of
magnitude.

\epsfxsize=7cm
\epsfysize=7.44cm
\begin{figure}[t]
	\epsffile{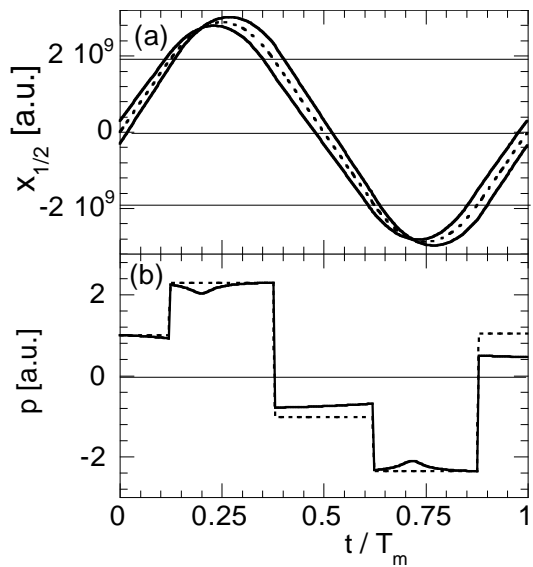}
	\caption{Solution of the equations of motion for two ions for one
	period $T$ with and without interaction. These plots should be
	compared to the results of the mapping (figure
	\ref{fig:ModTrajTypes}).
	Panel (a) plots the positions: the distance of the two ions
	(------) from the CM's trajectory (- - -) is multiplied by a
	factor of 300. The horizontal lines at $x_{1/2}=\pm1.89 \times
	10^9$ a.u. denote the beginning of the mirror regions.
	(b) shows the relative momentum between the ions with (------) and
	without (- - -) the interaction. The temporary slowing down due to
	the interaction is clearly visible.}
	\label{fig:EinUmlauf}
\end{figure}

Figure \ref{fig:EinUmlauf} we show the result of a trajectory
calculation for one period $T$ (\ref{eq:TimeForPeriod}) for the
initial values $x=-10^6$ a.u. and $p=1$ a.u. All other parameters were
set to the experiment's values of \tbl{tbl:Dimensions}, resulting in
$\alpha = 0.956$. 
Panel (a) shows the CM's position vs. time as a broken curve and the
two ions' coordinates. Their distance from the CM has been increased
by a factor of 300 for clarification. Case(i) of figure
\ref{fig:ModTrajTypes}(a) is reproduced during the time interval
$0.1T<t<0.4T$ and, symmetrically in the other mirror, during
$0.6T<t<0.9T$.
Panel (b) compares the relative momentum with (solid line) and without
(broken line) interaction: the "discontinuities" of the momentum kicks
are clearly seen, cf. figure \ref{fig:ModTrajTypes}. Without
interaction the momentum is reversed after each mirror and returns to
its initial value: $p'' = -p' = p$ after one full period. 
 
The main difference between the trajectory calculations and the
mapping description is that in our mapping approach the ions'
collisions are instantaneous and confined to the mirror. In the
trajectory calculations the ion--ion interaction takes place
anywhere in the trap on a time scale comparable to $T_m$ for small
relative momenta. This can be seen in figure \ref{fig:EinUmlauf}(b)
where the relative momentum deviates visibly from the non interacting
case for about half of the time spent in the mirror.

The trajectory shown in figure \ref{fig:EinUmlauf} confirms that the
processes of figures \ref{fig:TrajTypes} and \ref{fig:ModTrajTypes},
which we used to explain the mechanism, exist. In order to confirm
both the bunching condition and the dominance of the collisions in the
mirror, we created Poincar\'e sections from the trajectory
calculation, too. The stability of the synchronization shows up in the
overall elliptic or hyperbolic structure of the orbits. If the ions
are synchronized dominantly through delaying collisions in the mirror,
as we propose, the orbits will be (deformed) ellipses, whereas the
alternating ``bricking motion'' (cf. section \ref{sec:ExplainCentral})
will lead to a discrete structure.

\epsfxsize=8cm
\epsfysize=8.5cm
\begin{figure}[t]
	\epsffile{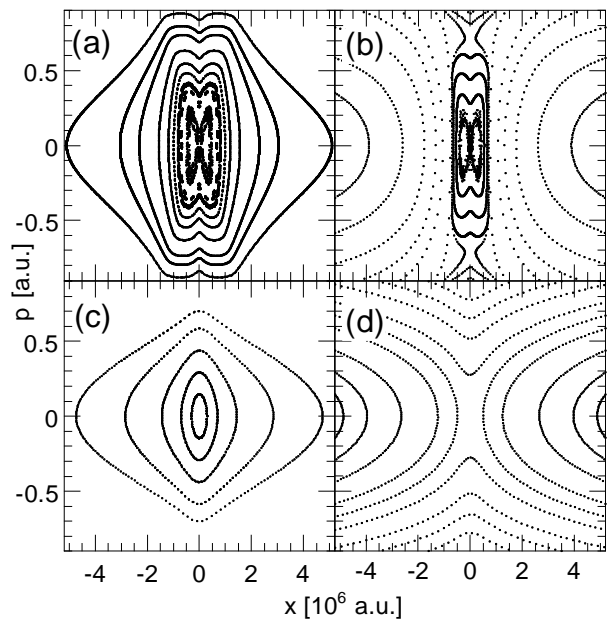}
	\caption{Poincar\'{e} sections from trajectory calculations for 
	the stable regime of $\alpha=0.956$ (a) and for the unstable 
	regime with $\alpha=1.046$ (b). The lower row is the result of the 
	repeated mapping $\mathcal{P}$ with a non constant $\tau_m$ from 
	\eq{eq:p3Delay} for the same values of $\alpha$ as (a) and (b): 
	(c) shows the stable and (d) the unstable regime.}
	\label{fig:PoincP3Num}
\end{figure}

The Poincar\'e sections from the trajectory calculation are shown in
figure \ref{fig:PoincP3Num} in panels (a) and (b). In the stable
regime, i.e., for $\alpha<1$, shown in (a), the orbits are deformed
ellipses: the ions orbit around each other. The strength of the
interaction, and therefore the local curvature of the orbit, varies
with the relative distance and the momentum, but it can still be
summarized by a time delay.

Only for small distances and momenta does the structure of the
Poincar\'e section become more complicated: the fixpoint at $(x,p) =
0$ is surrounded by a chaotic layer. When the ions are very close
together the effective delay changes so much between successive
collisions that the quasicontinuous orbits are broken up and
consecutive points are distributed randomly in the central region of
the Poincar\'{e} section.

The agreement between the Poincar\'e sections from the mapping and
from the trajectory calculations can be improved, when the constant
$\tau_m$ of figure \ref{fig:PoincConstTau} is replaced by a momentum
dependent delay. That the delay depends only on the relative momentum
is motivated by our ansatz that the ions reach the asymptotic region
again after the collisions. Then the delay does not depend upon the
initial and final distance, but only on the speed with which the ions
pass each other. In figure \ref{fig:PoincP3Num}(c) we show the result
of the mapping with the ansatz
\be{eq:p3Delay}
	\tau_m = \tau_m(p) = \frac{C}{\sqrt{|p/p_0|^3 + 1}}.
\ee
The functional form and the parameters --- $C = 3 \times 10^9$ a.u.
and $p_0 = 1$ a.u. --- were chosen to give orbits of a comparable
structure and size as the trajectory calculations of panel (a). 

In figure \ref{fig:PoincP3Num}(b) we increased $\alpha$ into the
unstable regime to $\alpha=1.046$. Then the trajectory calculation
becomes unstable, just as the mapping, except for a small island
around $(x,p)=0$. Even this island vanishes when $\alpha$ is increased
further. The stability condition is $\alpha<1$ as in the mapping, but
the transition is less abrupt than with the mapping approach.

Panel (d) plots the corresponding Poincar\'e section of the mapping
with the momentum dependent delay of equation \eq{eq:p3Delay}. Again,
as was the case with panels (a) and (c) in the bunching regime, the
overall structure is comparable to the trajectory calculation of (b).
The small stable island does not exist, though.

The Poincar\'e surfaces of section provide a further indication that
the mapping ansatz is an excellent representation of the full
trajectory dynamics with respect to both the dominant synchronization
mechanism and the stability condition.

\subsection{Non--linear mirror fields}

Above, we modelled the trap with a constant gradient mirror field
\eq{eq:trapPotential}. This allowed us to decouple the relative
coordinate from the CM motion. We now have to ensure that our results
and explanations are also valid when the mirror field is not linear in
the experiment.

Two conditions must be fulfilled for synchronization to occur in our
model (cf. sections \ref{sec:MirrorCollisions} and
\ref{sec:ExplainMirror}): (i) the total dispersion $\frac{\partial
T}{\partial p_0}$ of the trap has to be positive and (ii) the
interaction has to delay the ions in the mirror, so that the relative
momentum is not exactly reversed.

It is easy to see that the overall positive dispersion is not
restricted to constant mirror field configurations; any mirror field
that increases slower than harmonic will do. The mirror only has to be
``long'' enough so that it can compensate the negative dispersion of
the field free part of the trap.

If the mirror potential is not linear it nevertheless reflects the
individual momenta of two non interacting ions. The relative and the
CM motion can not be separated any more, but the evolution of the
relative coordinate --- and consequently the momentum reflection ---
can be described by an alternation of infinitesimally small kicks and
free evolutions.

The ions' interaction is then added into this succession of mappings
as a set of infinitesimal delays, each modifying the balance between
the adjacent kicks. The overall effect is the same as with the linear
slope: when the ions repel each other they leave the mirror with a
smaller distance and the second ion transfers energy onto the first.

Consequently the conditions for synchronization and the explanations
given remain unchanged for an arbitrary mirror field.

\subsection{Connection to previous explanations}

The approach presented here has several aspects in common with the one
recently given by Strasser \emph{etal} \cite{STR02}. In fact the
propagation matrices for the non interacting ions are identical. In
the Strasser model the composite mapping \eq{eq:MappNonInter} derived
from these matrices is cast into an effective reduced mass $m^*$ (see
equations (6) and (8) of reference \cite{STR02}). When the ions repel
each other, $m^*$ has to be negative for the relative motion to be
bound. This leads to the bunching condition of a positive dispersion
of the trap, i.e., $\frac{\partial T}{\partial p_0}>0$. Both the
trap's dispersion and the ions' repulsion are described in a mean
field ansatz; the Strasser model is therefore very robust with respect
to the exact form of the trap potential or the details of the ions'
interaction, but can not give an explanation of the underlying
microscopic dynamics.

Both the Strasser and our model require $\alpha<1$ for bunching, but
whereas in our approach \emph{any} delay $\tau_m>0$ is sufficient to
couple the ions, the Strasser model requires a minimum density in the
bunch, i.e., a minimum interaction strength, because it confines the
test ion to stay inside the bunch.

By comparing equation (10) of reference \cite{STR02} with our equation
\eq{eq:HalfP} we can relate the static bunch's parameters to an
effective time delay $\tilde{\tau}_m$:
%
\begin{equation}
	\tilde{\tau}_m = 
		\frac{Nq^2 T_m^3}{16 \pi m \epsilon_0 R_0^3} (1-\alpha)
\end{equation}
The bunch consists of $N$ ions of charge $q$ and mass $m$ and has a
radius $R_0$; $\epsilon_0$ is the dielectric constant.

From the phase space flow of the Poincar\'e sections we see that in
the diffusive regime the faster ions pass the pickup in the front of
the bunch. This was already observed in \cite{PED02b}. In the bunching
regime, though, this ordering is not simply inverted. The Poincar\'{e}
sections are symmetric with respect to the coordinate and the momentum
axis, consequently each of the ions is on average in front for half of
the time and has the higher momentum for half of the time, when the CM
passes the pickup.

\section{Summary and conclusions}

In this paper we have set up a model to describe the motion of two
ions in an "ion trap resonator", an ion trap, which essentially
consists of two spatially separated electrostatic mirrors. Our one
dimensional model describes the evolution of the ions' relative
coordinate and momentum with three simple mappings: one, each, for the
free evolution in the central field free region of the trap and in the
mirrors, while the third mapping describes the connecting kink between
these regions: here the relative coordinate is coupled to the center
of mass motion. The coupling gives rise to an instantaneous
compressing kick in the relative momentum. The interaction between the
two ions is summarized by a time delay, which shortens or lengthens
the time for the free evolution either in the mirror or in the central
part.

The three mappings allow us to perform a linear stability analysis of
the composite mapping for the whole trap. Without delay, i.e., without
interaction, the mapping model reproduces the behavior of two
independent ions; with the delay added we identify the criteria for
the stable synchronized motion: the trap has to be operated in a
regime where the period of the ions in the trap increases with their
energy \emph{and} the ions have to be delayed by their repulsive
interaction, when they cross their paths in the mirror. This confirms
the experimental findings that the dispersion of the trap has to be
positive.

Based on these simple building blocks we are able to describe, how the
interaction modifies and couples the motion of the two ions in the
bunching regime: without the interaction ions with different energies
separate linearly in time due to their different periods in the trap.
A time delay in the mirror, which describes the repulsive interaction,
now modifies the balance of the coupling between the relative and the
CM motion at the kink, so that during each pass through the mirror
momentum is transferred from the second onto the first ion. This
additional energy increases the first ion's period and lets it fall
back against the other one until they have exchanged their places. Now
the energy is shuffled back onto the other ion, until they exchange
their position again and the circle is completed: in the stable region
the ions oscillate around each other, constantly transferring energy
back and forth between them.

Trajectory calculations confirmed that the ions are synchronized
dominantly by the microscopic process of collisions in the mirror that
we had derived from the mappings together with an instantaneous
interaction. This continuous description reproduces the stability
conditions, too. We also verified that our model is valid for
arbitrary forms of the mirror potential.

Our model of two ions can also be applied to the dynamics of a test
ion inside a bunch. The microscopic picture of the ions' motion, which
was developed in this paper, then allows us to identify the important
regions of the trap and to explain how they shape the bunch. This will
be the subject of a forthcoming publication \cite{GEY02}.

\begin{acknowledgments}

We thank Daniel Strasser and Daniel Zajfman for constructive
discussions and further explanations of the experiment.

This research was funded by the Israel Science Foundation.

\end{acknowledgments}


\end{document}